\def\rfr#1{eq. (\ref{#1})}

\def\dert#1#2{\frac{{{d}}{#1}}{{{d}}{#2}}}              

\def\virg#1{``#1''}

\def\eqi{\begin{equation}}
\def\eqf{\end{equation}}
\def\eqia{\begin{eqnarray}}
\def\eqfa{\end{eqnarray}}
\def\rp#1#2{{#1\over#2}} \def\lb#1{\label{#1}}

\def\bds#1{\boldsymbol{#1}}

\documentclass[11pt]{article}

\usepackage{slashbox}
\usepackage{url}
\usepackage{amsmath,amsthm,amscd,amssymb}
\usepackage{latexsym,wasysym}
\usepackage{graphicx,epsfig}
\RequirePackage{color}

\begin{document}

	
\noindent{\bf \LARGE{An Empirical  Explanation of  the Anomalous Increases in the Astronomical Unit and  the Lunar Eccentricity }}
\\
\\
\\
{L. Iorio$^{\ast}$}\\
{ $^{\ast}$Ministero dell'Istruzione, dell'Universit\`{a} e della Ricerca (M.I.U.R.)-Istruzione\\ Fellow of the Royal Astronomical Society (F.R.A.S.)\\
International Institute for Theoretical Physics and
High Mathematics Einstein-Galilei\\
 Permanent address: Viale Unit$\grave{a}$ di Italia 68
70125 Bari (BA), Italy.  \\ e-mail: lorenzo.iorio@libero.it}\\
{\it Received 2011 May 5; accepted 2011 June 17; published 2011 July 25}

\begin{abstract}
   {The subject of this paper is the empirically determined anomalous secular increases of the astronomical unit, of the order of some cm yr$^{-1}$, and of the eccentricity of the lunar orbit, of the order of $10^{-12}$ yr$^{-1}$.}
   {The aim is to find an empirical explanation of both the anomalies, as far as their orders of magnitude are concerned.}
   {The methods employed are working out perturbatively with the Gauss equations  the secular effects on the semi-major axis $a$ and the eccentricity $e$ of a test particle orbiting a central body acted upon by a small anomalous radial acceleration $A$ proportional to the radial velocity $v_r$ of the particle-body relative motion.}
   {The results show that  non-vanishing secular variations $\left\langle\dot a\right\rangle$ and $\left\langle\dot e\right\rangle$  occur. If the magnitude of the coefficient of proportionality of the extra-acceleration is of the same order of magnitude of the Hubble parameter $H_0=7.47\times 10^{-11}$ yr$^{-1}$ at the present epoch, they are able to explain both the astrometric anomalies without contradicting other existing observational determinations for the Moon and the other planets of the solar system. }
   {Finally, it is concluded that the extra-acceleration might be of cosmological origin, provided that the relative radial particle-body motion is accounted for in addition to that due to the cosmological expansion only. Anyway, further data analyses should confirm or disproof the existence of both  the astrometric anomalies as genuine physical phenomena.}
\end{abstract}


\section{Introduction}
Recently, {the main features of the} anomalous secular increases of both the astronomical unit and  the eccentricity $e$ of the lunar orbit {have} been {reviewed} \cite{And010}. While the first effect, obtained by several independent researchers \cite{Kra,Sta,Pit,Pit2,And010}, should be of the order of a few cm yr$^{-1}$, the second one {\cite{ecce1,ecce2}} amounts to  $\dot e = (9\pm 3)\times 10^{-12}$ yr$^{-1}${, according to the latest data analysis \cite{ded}}.

 {Such phenomena attracted the attention of various  scientists dealing with them in different contexts \cite{Ni,Kope1,newlamm,revi,Kope2,Li1,Li2,Gold,Spelio,Zha,Sha,Lamm11,Zha2}.} {Thus, s}everal more or less sound attempts to find{, or to rule out, possible}  explanations \cite{Kra,Iorio05,Mash0,Ostv,Mash,Klo,Nord,Lamm,Verb,Lamm2,Amin,Araki1,Jap1,Jap2,cinesi,Lamm3,Brumm,Nya,Bel,Araki2,And010,Raso,Iorio011,Ara011} for both the anomalies  were  proposed so far, both in terms of standard known  gravitational physical phenomena and of long-range modified models of gravity.

Here we propose an empirical formula which is able to accommodate both the anomalies, at least as far as their orders of magnitude are concerned.
\section{An anomalous acceleration proportional to the radial velocity of the test particle}
Let us assume that, in addition to the usual Newtonian inverse-square law for the gravitational acceleration  imparted to a test particle by a central body orbited by it, there is also a small radial extra-acceleration of the form
\eqi A_{\rm pert}= k H_0 v_r.\lb{hubacc}\eqf In it $k$ is a positive numerical parameter of the order of unity to be determined from the observations, $H_0=(73.8\pm 2.4)$ km s$^{-1}$ Mpc$^{-1}=(7.47\pm 0.24)\times 10^{-11}$ yr$^{-1}$ \cite{hubb} is the Hubble parameter at the present epoch, defined in terms of the time-varying cosmological scaling factor $S(t)$ as $H_0\doteq \left.\dot S/S\right|_0$, and $v_r$ is the component of the velocity vector $\bds v$ of the test particle's proper motion about the central body along the common radial direction.
The radial velocity for a Keplerian ellipse is \cite{Brou}
\eqi v_r=\rp{nae\sin f}{\sqrt{1-e^2}},\eqf where $n$ is the Keplerian mean motion, $a$ is the semi-major axis, and $f$ is the true anomaly reckoning the instantaneous position of the test particle along its orbit: $v_r$ vanishes for circular orbits.

The consequences of \rfr{hubacc} on the trajectory of the particle can be straightforwardly worked out with the standard  Gauss equations for the variation of the Keplerian orbital elements \cite{Brou} which are valid for any kind of perturbing acceleration, whatever its physical origin may be.
For the semi-major axis and the eccentricity they are
\begin{equation}
\left\{
\begin{array}{lll}
\dert a t & = & \rp{2}{n\sqrt{1-e^2}} \left[e A_R\sin f +A_{T}\left(\rp{p}{r}\right)\right],\\   \\
\dert e t  & = & \rp{\sqrt{1-e^2}}{na}\left\{A_R\sin f + A_{T}\left[\cos f + \rp{1}{e}\left(1 - \rp{r}{a}\right)\right]\right\}.
\end{array}\lb{Gauss}
\right.
\end{equation}
In \rfr{Gauss} $p\doteq a(1-e^2)$ is the semi-latus rectum, and $A_R$ and $A_T$ are the radial and transverse components of the disturbing acceleration, respectively: in our case, \rfr{hubacc}  is entirely radial. In a typical first-order perturbative\footnote{Indeed, it can be easily inferred that \rfr{hubacc} is of the order of $10^{-15}$ m s$^{-2}$ for the Earth's motion around the Sun, while its Newtonian solar monopole term is as large as $10^{-3}$ m s$^{-2}$. The same holds for the Earth-Moon system as well. Indeed, \rfr{hubacc} yields about $10^{-16}$ m s$^{-2}$ for the lunar geocentric orbit, while the Newtonian monopole acceleration due to the Earth is of the order of $10^{-3}$ m s$^{-2}$.} calculation like in the present case, the right-hand-sides of \rfr{Gauss} have to be computed onto the unperturbed Keplerian ellipse, characterized by
\eqi r=\rp{p}{1+e\cos f},\eqf and integrated over one orbital period by means of
\eqi dt = \rp{1}{n}\left(\rp{r}{a}\right)^2 \rp{1}{\sqrt{1-e^2}} df.\eqf
It turns out that both the semi-major axis $a$ and the eccentricity $e$ of the test particle's orbit secularly increase according to
\begin{equation}
\left\{
\begin{array}{lll}
\left\langle \dot a \right\rangle & = & 2ka H_0 \left(1-\sqrt{1-e^2}\right) , \\ \\
\left\langle \dot e \right\rangle & = & kH_0 \rp{\left(1-e^2\right)\left(1-\sqrt{1-e^2}\right)}{e}.
\end{array}\lb{syste}
\right.
\end{equation}
The formulas in \rfr{syste}, which were obtained by taking an average over a full orbital revolution, are exact to all order in $e$.

Since $e_{\rm Moon}=0.0647$, it turns out that \rfr{syste} is able to reproduce the measured anomalous increase of the lunar orbit for $2.5\lesssim k\lesssim 5$. Moreover,  for such  values of $k$ \rfr{syste} yields an increase of the lunar semi-major axis of just $0.3-0.6$ mm yr$^{-1}$. It is, at present, undetectable, in agreement with the fact that, actually, no anomalous secular variations pertaining such an orbital element of the lunar orbit have been detected so far. If we assume the terrestrial semi-major axis\footnote{Actually, the astronomical unit is neither the semi-major axis of the Earth's orbit nor its average distance from the Sun $\left\langle r \right\rangle=a(1+e^2/2)$ \cite{Sta}. On the other hand, it is $a_{\oplus}=1.00000018$ au (http://ssd.jpl.nasa.gov/txt/p$\_$elem$\_$t2.txt).} $a_{\oplus}=1.5\times 10^{13}$ cm  as an \textit{approximate} measure of the astronomical unit and consider that $e_{\oplus}=0.0167$, \rfr{syste} and the previous values of $k$ yield a secular increase of just a few cm yr$^{-1}$. Also in this case, it can be concluded that \rfr{syste}, if applied to other situations for which accurate data exist, does not yield results in contrast with empirical determinations for $a$ and $e$. Indeed, for the eccentricity of the Earth \rfr{syste}, with $2.5\lesssim k\lesssim 5$, yields $\left\langle\dot e \right\rangle=(1.7-3.4)\times 10^{-12}$ yr$^{-1}$. Actually,   such an anomalous effect cannot be detectable since, according to Table 3 of Ref.~\cite{pita}, the present-day formal, statistical accuracy in determining $e$ from the observations amounts just to $3.6\times 10^{-12}$; it is well known that the realistic uncertainty can be up to one order of magnitude larger. Similar considerations hold for the other planets.
\section{Conclusions}
Here we do not intend to speculate too much about possible viable physical mechanisms yielding the extra-acceleration of\footnote{See Ref.~\cite{Raso} and his quasi-Newtonian dynamics, intermediate between Newtonian and modified Newtonian dynamics.} \rfr{hubacc}.

It might be argued  that, reasoning within a cosmological framework, the Hubble law may give \rfr{hubacc} for $k=1$ if the proper motion of the particle about the central mass is taken into account in addition to its purely cosmological recession which, instead, yields the well-known local\footnote{For a recent review of the influence of global cosmological expansion on the local dynamics and kinematics, see Ref.~\cite{revi}.} extra-acceleration of tidal type \cite{Cooper,Mash,Klio1} \eqi A_{\rm cosmol} = -q_0 H^2_0 r,\eqf where $q_0\doteq -\left(\ddot{S}/S\right)_0 H_0^{-2}$ is the deceleration parameter at the present epoch.

On the other hand, our empirical results,  which are not in contrast with other observational determinations for the Moon and the other planets of the solar system, may be  simply interpreted, in a purely phenomenological way, in terms of a radial extra-acceleration proportional to the radial component $v_r$ of the proper velocity of the test particle about its primary through a coefficient having the dimensions of T$^{-1}$ and a magnitude close to that of $H_0$: its physical origin should not necessarily be of cosmological origin.

{Finally, we want to spend some words  about the nature of the anomalies considered. As it was shown, the anomalous increase of the astronomical unit has attracted the attention of several researchers so far. It was determined as a solve-for parameter by using different ephemerides which neither use the same dynamical force models nor the same observational records. Further processing of more extended data sets, with more accurate dynamical modeling, will be useful in shedding further light on such an anomaly. The authors of Ref.~\cite{And010} conclude their review of the  anomalous variation of the astronomical unit by writing at pag. 194: \virg{If the reported increase holds up under further scrutiny and additional data
analysis, it is indeed anomalous. Meanwhile it is prudent to remain skeptical of any real
increase. In our opinion the anomalistic increase lies somewhere in the interval zero to
20 cm yr$^{-1}$ , with a low probability that the reported increase is a statistical false alarm.} On the other hand, it must be remarked that a clear definition for the change of the astronomical unit is still lacking since some researchers believe that the astronomical unit  is a redundant unit, like to the gravitational parameter $GM$ of the Sun, which should, instead, be empirically determined from data processing as a solve-for parameter \cite{Klio,Fie,Jou}. It is likely that at the IAU meeting in 2012 a fixed numerical value for it will be adopted\footnote{W. M. Folkner, personal communication to the author, February 2011.}. Anyway, this would have the effect of just shifting the detected anomaly to another physical quantity.  Concerning the Moon's orbit, the first report of the lunar eccentricity anomaly dates back to 2001 \cite{ecce1}; such a phenomenon is still here \cite{ecce2,ded}, despite the increasing accuracy in LLR observations and modeling occurred in the last decade due to the steady efforts of a wide community of researchers engaged in LLR science and technology. On the other hand, it is not certainly unreasonable to expect  that further modeling of classical effects occurring in the lunar interior may finally be able to explain the observed anomaly. Anyway, until it will actually happen, looking for alternative explanations remains a task worth being pursued.}
%


\begin{thebibliography}{}

\bibitem{And010}
Anderson J. D., Nieto M. M., 2010, Astrometric solar-system anomalies. In:  Klioner S.A.,   Seidelmann P.K.,  Soffel M.H.,  (eds.)  Relativity in Fundamental Astronomy: Dynamics,
Reference Frames, and Data Analysis,
Proceedings IAU Symposium No. 261, Cambridge University Press, Cambridge, 2010, pp. 189-197

\bibitem{Kra}
Krasinsky G. A.,  Brumberg V. A., 2004,  Celest. Mech. and Dyn.
Astron, 90,  267

\bibitem{Sta}
Standish E. M., 2005, The Astronomical Unit now. In:  Kurtz D. W. (ed.),
Transits of
Venus: New Views of the Solar System and Galaxy, Proceedings of the IAU Colloquium No. 196, Cambridge University Press, Cambridge, 2005, pp. 163-179

\bibitem{Pit}
Pitjeva E. V., Standish E. M., 2005, The Astronomical Unit now. In:  Kurtz D. W. (ed.),
Transits of
Venus: New Views of the Solar System and Galaxy, Proceedings of the IAU Colloquium No. 196, Cambridge University Press, Cambridge, 2005, p. 177

\bibitem{Pit2}
E. V. Pitjeva, E. V., 2008, private communication to Noerdlinger P.

{
\bibitem{ecce1}
Williams J. G., Boggs D. H., Yoder C. F., Ratcliff J. T.,
Dickey J. O., 2001, J. Geophys. Res., 106, 27933
}

{
\bibitem{ecce2}
Williams J. G., Dickey J. O., 2003, Lunar Geophysics,
Geodesy, and Dynamics. In: Noomen
R., Klosko S., Noll C., Pearlman M. (eds.) Proceedings
of the 13th International Workshop on
Laser Ranging \virg{Toward Millimeter Accuracy}, Wshington D.C., October 7-11 2002. NASA/CP-2003-212248, pp. 75-86.
http://cddis.nasa.gov/lw13/docs/papers/sci$\_$williams$\_$1m.pdf
}

\bibitem{ded}
Williams J. G., Boggs D. H., 2009, Lunar Core and Mantle. What Does LLR See? In:
Schilliak S. (eds.) Proceedings of the 16th International Workshop on Laser Ranging \virg{SLR-the next generation}, Pozna\'{n}, October 13-17 2008.
http://cddis.gsfc.nasa.gov/lw16/docs/papers/sci$\_$1$\_$Williams$\_$p.pdf

{
\bibitem{Ni}
Ni W.-T., 2005, Int. J. Mod. Phys. D, 14, 901
}

{
\bibitem{Kope1}
Kopeikin S. M., 2007, Relativistic Reference Frames for Astrometry and Navigation in the Solar System. In: Belbruno E. (ed.)
NEW TRENDS IN ASTRODYNAMICS AND APPLICATIONS III. AIP Conference Proceedings, Volume 886, pp. 268-283
}

{
\bibitem{newlamm}
L\"{a}mmerzahl C., Dittus H., 2007, Int. J. Mod. Phys. D, 16, 2455
}

\bibitem{revi}
Carrera M., Giulini D., 2010, Rev. Mod. Phys., 82, 169

{
\bibitem{Kope2}
Kopeikin S. M., 2010, Beyond the standard IAU framework. In:  Klioner S. A.,  Seidelmann P. K.,  Soffel M. H. (eds.) Relativity in Fundamental Astronomy: Dynamics, Reference Frames, and Data Analysis, Proceedings of the International Astronomical Union, IAU Symposium, Volume 261, Cambridge University Press, Cambridge, pp. 7-15
}

{
\bibitem{Li1}
Li X., Chang Z., Li M., 2010, arXiv:1001.0066
}

{
\bibitem{Li2}
Li X., Chang Z., Mo X., 2010, arXiv:1001.2667
}

{
\bibitem{Gold}
Goldhaber A. S., Nieto M. M., 2010, Rev. Mod. Phys., 82, 939
}



{
\bibitem{Spelio}
Speliotopoulos A. D., 2010, Gen. Relativ. Gravit., 42, 1537
}

{
\bibitem{Zha}
Zhang W. J., Li Z. B., Lei Y., 2010, Chinese Sci. Bull., 55, 4010
}


{
\bibitem{Sha}
Sharma B. K., 2011, Earth, Moon, and Planets, 108, 15
}

{
\bibitem{Lamm11}
L\"{a}mmerzahl C., 2011, Testing Basic Laws of Gravitation-Are Our Postulates on Dynamics and Gravitation Supported by Experimental Evidence? In: Blanchet L., Spallicci A.,  Whiting B. (eds.)
Mass and Motion in General Relativity
Fundamental Theories of Physics, Volume 162, Springer, Berlin, pp. 25-65
}

{
\bibitem{Zha2}
Zhang W. J., Kelley N., 2011, Advanced Sci. Lett., 4, 574
}

\bibitem{Iorio05}
Iorio L., 2005, J.  Cosmol. and Astropart. Phys., 09, 006

{
\bibitem{Mash0}
Mashhoon B., Singh D., 2006, Phys. Rev. D., 74,  124006
}

\bibitem{Ostv}
{\O}stvang D., 2007, Gravit. Cosmol., 13, 1

\bibitem{Mash}
Mashhoon B., Mobed N., Singh D., 2007, Class. Quantum Gravit., 24, 5031

\bibitem{Klo}
Khokhlov D. L., 2007, arXiv:0710.5862

\bibitem{Nord}
Noerdlinger P., 2008, arXiv:0801.3807

{
\bibitem{Lamm}
L\"{a}mmerzahl C.,  Preuss O., Dittus H., 2008, Is the Physics Within the Solar System Really Understood? In: Dittus H., L\"{a}mmerzahl C., Turyshev S. G. (eds.) Lasers, Clocks and Drag-Free Control
Exploration of Relativistic Gravity in Space, Astrophysics and Space Science Library
Volume 349, Springer, Berlin, pp. 75-101
}

{
\bibitem{Verb}	
Verbiest J. P. W., Bailes M., van Straten W., Hobbs G. B., Edwards R. T.,
Manchester R. N., Bhat N. D. R., Sarkissian J. M., Jacoby B. A., Kulkarni S. R., 2008, Astrophys. J., 679,  675
}

{
\bibitem{Lamm2}
L\"{a}mmerzahl C., 2008, Eur. Phys. J. Special Topics, 163, 255
}

{
\bibitem{Amin}
Amin M. Y., 2009, arXiv:0912.2443
}

\bibitem{Araki1}
Arakida H., 2009, New Astron., 14,  264

\bibitem{Jap1}	
Miura T., Arakida H., Kasai M., Kuramata S., 2009, Publ. Astron. Soc. Jp., 61, 1247

\bibitem{Jap2}	
Ito Y., 2009, Publ. Astron. Soc. Jp., 61, 1373

\bibitem{cinesi}	
Li X., Chang Z., 2009, arXiv:0911.1890

{
\bibitem{Lamm3}
L\"{a}mmerzahl C., 2009, Space Sci. Rev., 148,  501
}

{
\bibitem{Brumm}
Brumberg V.A., 2010, Celest. Mech. Dyn. Astron., 106, 209
}

\bibitem{Nya}
Nyambuya G. G, 2010, Mon. Not. Roy. Astron. Soc, 403, 1381

{
\bibitem{Bel}
Bel Ll., 2010, arXiv:1005.5442
}

\bibitem{Araki2}
Arakida H., 2010, Adv. Sp. Res., 45,  1007

\bibitem{Raso}
Rasor N. S., 2010, Physics Essays, 23, 383

\bibitem{Iorio011}
Iorio L., 2011, Mon. Not.  Roy. Astron. Soc., doi:10.1111/j.1365-2966.2011.18777.x
{
\bibitem{Ara011}
Arakida H., 2011, Gen. Relativ. Gravit., doi:10.1007/s10714-011-1170-1
}

\bibitem{hubb}	
Riess A. G.,  Macri L., Casertano S., Lampeit H., Ferguson H. C.,  Filippenko A. V.,  Jha S. W.,  Li W., Chornock R., Silverman J. M. , 2011, Astrophys. J., 730, 119

\bibitem{Brou}
Brouwer D., Clemence G. M., 1961, Methods of celestial mechanics, Academic Press  (New York)



\bibitem{Cooper}
Cooperstock F. I., Faraoni V., Vollick D. N., 1998,
Astrophys. J., 503, 61

{
\bibitem{Klio1}
Klioner S. A., Soffel M. M., 2005, Refining the Relativistic Model for Gaia: Cosmological Effects in the BCRS In: Turon C.,  O'Flaherty K. S.,
Perryman M. A. C. (eds.), Proceedings of the Symposium
\virg{The Three-Dimensional Universe with Gaia}, October 4-7 2004, Paris. ESA SP-576, pp. 305-308
}

{
\bibitem{Klio}
Klioner S. A., 2008, Astron. Astrophys., 478, 951
}

{
\bibitem{Fie}	
Fienga A., Manche H., Kuchynka P., Laskar J., Gastineau M., 2010, arXiv:1011.4419
}

{
\bibitem{Jou}
Capitaine N., Guinot B., Klioner S., 2010, Proposal for the re-definition of the astronomical unit of length through a fixed relation
to the SI metre.  Rencontres de l'Observatoire
Journ\'{e}es 2010 \virg{Syst\`{e}mes de r\'{e}f\'{e}rence spatio-temporels}. New challenges
for reference systems
and numerical standards
in astronomy. Observatoire de Paris
Ecole Normale Sup\'{e}rieure
20-22 September 2010. p. 6
}

\bibitem{pita}
Pitjeva E. V., 2008, Use of optical and radio astrometric
observations of planets, satellites and
spacecraft for ephemeris astronomy. In: Jin W. J.,  Platais I.,  Perryman M. A. C. (eds.) A Giant Step: from Milli- to Micro-arcsecond Astrometry
Proceedings IAU Symposium No. 248. Cambridge University Press, Cambridge, pp. 20-22.

\end{thebibliography}
\end{document}